\documentclass[12pt,preprint]{aastex}

\slugcomment{}

\shorttitle{ The compact circumstellar material around 
OH~231.8+4.2}
\shortauthors{Matsuura et al.}

\begin{document}


\title{The compact circumstellar material around 
OH~231.8+4.2\footnote{
Based on observations made with the VLT and
the VLTI (Project numbers 065.L-0395, 072.D-0766 and 074.D-0405)} }


\author{M.~Matsuura\altaffilmark{2,3,4}, 
O.~Chesneau\altaffilmark{5},
A.A.~Zijlstra\altaffilmark{2},
W.~Jaffe\altaffilmark{6},
L.B.F.M.~Waters\altaffilmark{7,8},
J.~Yates\altaffilmark{9},
E.~Lagadec\altaffilmark{2},
T.~Gledhill\altaffilmark{10},
S.~Etoka\altaffilmark{2}, 
and A.M.S.~Richards\altaffilmark{11}
}
%\affil{Astronomy Department, University of California,
%    Berkeley, CA 94720}

%\author{C. D. Biemesderfer\altaffilmark{4,5}}
%\affil{National Optical Astronomy Observatories, Tucson, AZ 85719}
%\email{aastex-help@aas.org}
%\and
%\author{R. J. Hanisch\altaffilmark{5}}
%\affil{Space Telescope Science Institute, Baltimore, MD 21218}


\altaffiltext{2}{School of Physics and Astronomy, University of Manchester,
P.O. Box 88, Manchester M60 1QD, UK}
\altaffiltext{3}{School of Mathematics and Physics, 
Queen's University of Belfast, Belfast BT7 1NN, UK}
\altaffiltext{4}{National Astronomical Observatory of Japan, Osawa 2-21-1, 
Mitaka, Tokyo 181-8588, Japan}
\altaffiltext{5}{Observatoire de la C\^{o}te d'Azur, 
Avenue Copernic, F-06130, Grasse, France}
\altaffiltext{6}{Leiden Observatory, P.B. 9513, Leiden 2300 RA, The Netherlands}
\altaffiltext{7}{Astronomical Institute `Anton Pannekoek', University of Amsterdam,
Kruislaan 403, 1098 SJ, Amsterdam, The Netherlands}
\altaffiltext{8}{Instituut voor Sterrenkunde, Katholieke Universiteit Leuven, 
Celestijnenlaan 200B, 3001 Heverlee, Belgium}
\altaffiltext{9}{University College London, Gower street, London, WC1E 6BT, 
 UK}
\altaffiltext{10}{Centre for Astrophysics Research, 
Science and Technology Research Institute, University of Hertfordshire, College Lane, 
Hatfield, Hertfordshire AL10 9AB, UK}
\altaffiltext{11}{Jodrell Bank Observatory, University of Manchester, 
Jodrell Bank, Cheshire SK11 9DL, UK}


\begin{abstract}
 We have observed the bipolar post-AGB candidate OH 231.8+4.2, using
the mid-infrared interferometer MIDI and the infrared camera with the
adaptive optics system NACO on the Very Large Telescope. An unresolved
core ($<$200~mas in FWHM) is found at the center of the OH\,231.8+4.2
in the 3.8~$\mu$m image.  This compact source is resolved with the
interferometer. 
We used two 8-meter telescopes with four different
baselines, which cover projected baseline lengths from 62 to 47 meters,
and projected position angles from 112 to 131 degrees that are almost
perpendicular to the bipolar outflow.  Fringes from 8 to 9~$\mu$m and
from 12 to 13.5~$\mu$m were clearly detected, whilst the strong
silicate self-absorption allows only marginal detection of
visibilities between 9 and 12~$\mu$m.  The fringes from the four
baselines consistently show the presence of a compact circumstellar
object with an inner radius of 30--40~mas, which is equivalent to
40--50 AU at 1.3~kpc.  This clearly shows that the mid-infrared
compact source is not the central star (3~AU) but circumstellar
material.  The measured size of the circumstellar material is
consistent with the size of such disks calculated by hydrodynamic
models, implying the circumstellar material may have a disk
configuration. 

\end{abstract}

\keywords{    stars: AGB -- post-AGB 
    stars: mass loss --
    (ISM:) dust, extinction
    ISM: jets and outflows --
    Infrared: stars ---             }

%________________________________________________________________

\maketitle
\section{Introduction}
 Low and medium mass stars ($\sim$1--8 $M_{\sun}$ on the main
sequence) experience an intensive mass-loss phase during the Asymptotic
Giant Branch (AGB) phase.  Typically, the AGB wind is spherically symmetric.
However, during the next evolutionary stages, i.e., the post-AGB
phase and the planetary nebula (PN) phase, a high fraction of stars
show asymmetric shapes in their circumstellar envelopes, such as
elliptical and bipolar.  One of the hypotheses about the formation of the
bipolar shape invokes a binary disk scenario \citep{Balick02,
VanWinckel03}.  Part of the material lost during the intensive AGB
mass-loss wind is trapped in the binary system, and a circumbinary disk is 
formed in the plane of the binary orbit.  
The disk restricts the direction of the low
density but high velocity post-AGB and PN wind in the equatorial
plane, and focuses the wind towards two poles.  The size of the binary
disk will be small \citep[less than 100 AU (80~mas for our
target);][]{Mastro99}, and so requires interferometric observations to
be resolved.

\object{OH 231.8+4.2} (IRAS~07399$-$1435; RA 07h42m16.83s Dec
$-$14d42m52.1s; hereafter OH~231) is one of the well studied post-AGB
candidates.  TiO bands are detected from the central region,
suggesting the central star exhibits a M9 spectral type  \citep{Cohen81}.  
\citet{SanchezContreras04} claimed 
the presence of a spectroscopic binary from optical spectra, because in addition to TiO
and VO bands from the M-type star, Balmer lines and continuum excess
are detected.  OH~231 is probably located in the open cluster, M~46
\citep{Jura85}, thus the distance is relatively well determined (1.3
kpc).

The outflow is strongly bipolar, and bubbles and
shocked regions are found in the outflow \citep[e.g.][]{Bujarrabal02}.
 L-, N- and Q-band seeing-limited images show an unresolved core at
 the center of this object \citep{Kastner92, Jura02}.  The infrared
 color of this compact source is extremely red, and it is believed
 to be a dusty disk \citep{Jura02}.  The velocity structure of
 SiO masers also suggest the presence of a rotating disk around this
 compact source \citep{SanchezContreras02}. 
 The OH masers appear to be associated with an expanding torus 
 \citep{Zijlstra01}.  

In this paper 
 we present both high resolution infrared (IR) images and 
 and mid-IR interferometeric visibilities of the
 central compact source, so as to resolve the
 compact source at the center and so as  to determine if this source is a disk.

\section{ Observations and Analysis}

 OH\,231 was observed with the MID-infrared Interferometric instrument
\citep[MIDI;][]{Leinert03} on the Very Large Telescopes (VLTs) on 2005
March 2nd (UT), using the telescopes Melipal and Yepun.
The observing log is summarized in Table~\ref{table-log}.  
From these observing runs, we also obtained acquisition images (single dish) 
with the adaptive optics (AO) system, spectro-photometric data, and 
visibilities (correlated fluxes) for each baseline.

The data reduction software packages
MIA and EWS
\citep{Jaffe04, Chesneau05a} were used to reduce the spectra and
visibilities.  To estimate errors in the visibility data 
the MIA visibilities were extracted with three
different thresholds of noise level, and the difference in the
visibilities between these thresholds are counted as errors.  
The difference between visibilities due to the choice of the calibration
data was counted as a systematic error, and added to the errors derived above.
The EWS visibilities are slightly higher than the MIA ones but well within 
error bars.

For spectro-photometry, we used \cite{Cohen99}'s templates of HD~139127
(K4.5III) for HD~50778 (K4III) and of HD~180711 (G9III) for HD~61935
(G9III), and scaled the flux to IRAS 12~$\mu$m measurements.  The
spectro-photometric error bars were estimated from eight calibrated
spectra recorded.

We observed OH~231 with the AO system and the
infrared camera NACO \citep{Rousset03, Lenzen03} on theVLT 
on the 2004 March 6th (UT).  
%The weather was clear.
The exposure time was 7.5~min for NB2.12 and 4.8~min (positive image
only) for L'-band, respectively.  Both filters measure
continuum emission.  The NB2.12 filter ($\lambda_{\rm c}$=2.122~$\mu$m
and $\Delta \lambda$=0.022~$\mu$m) is designed to detect H$_2$ 1--0 S(1)
line, however, our spectroscopic observations using ISAAC/VLT show no
detection of this line, and the NB2.12 band measured continuum emission,
probably scattered light.  We used the S27 and L27 cameras (a pixel
scale of 27~mas).  The central wavelength is 3.80~$\mu$m and the width
is 0.62~$\mu$m for L'-band.  The wave-front sensor was used in visible
light provided by an AO reference star ($V$=14.2~mag) approximately
35\,arcsec away from the central compact object, giving a Strehl
ratio of about 30~\% at 2.12~$\mu$m.  An ND filter was used for
the L'-band observations, which reduces the energy by 1.5--2~\%.  The
jitterring technique was used to minimize the effect of hot pixels and
the sky background was estimated from a medium of jitterred frames
with different positions. A chopping technique was used during the L'-band
observations with a throw of 10 arcsec to the east and the
west.

MERLIN Phase-referenced observations of the 1667~MHz OH maser line
have been obtained on the 2005 April 25th, using a velocity
resolution of 0.7~km\,s$^{-1}$. 
The observations covered
the velocity range, 
from $v_{\rm LSR}$=$-$20 to +80\,km\,s$^{-1}$.
The angular resolution is about 0.2 arcsec. The data were reduced
using AIPS. Velocity maps were obtained by calculating moment maps.
The systemic velocity is about 35~km\,s$^{-1}$ \citep{Zijlstra01}.

\section{Data description and interpretation}

\begin{figure*}
\centering
%\resizebox{\hsize}{!}{\includegraphics*{2_12_fig_col.eps}}
\resizebox{\hsize}{!}{\includegraphics*{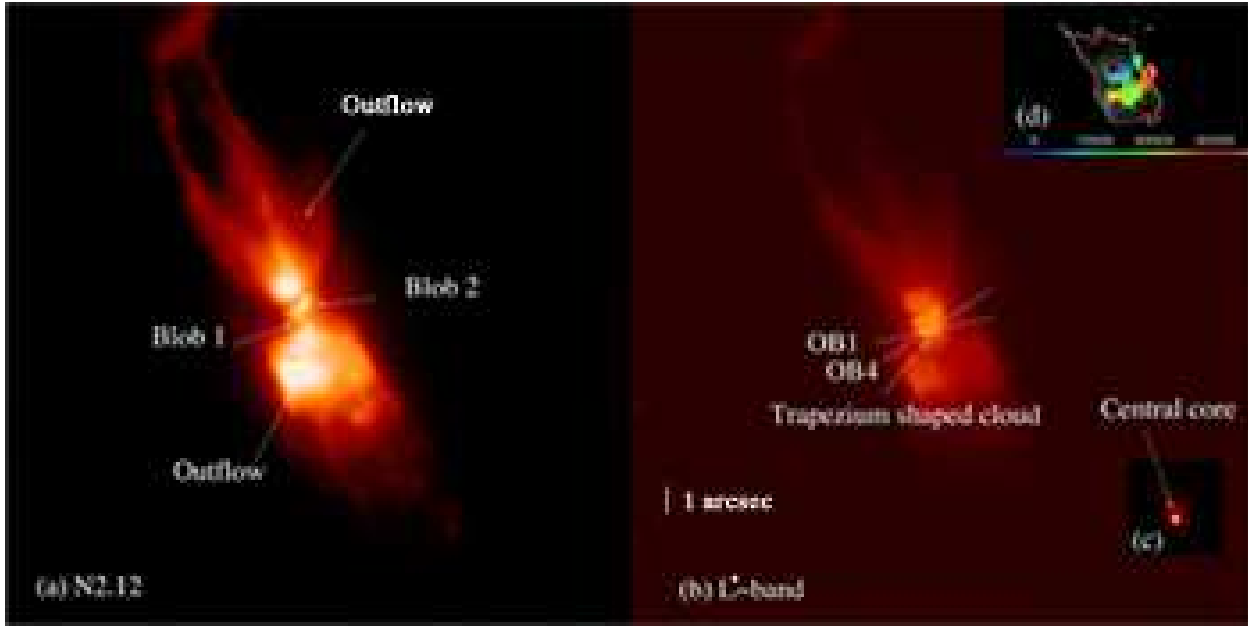}}
\caption{Near infrared adaptive optics images (a, b and c) of OH 231.
Lines in image (b) show the approximate baseline position angle of OB1
and OB4 for MIDI observations.  Insert (c) shows the brightest region of
L'-band image in different color scale so as to clearly show the
unresolved `central' object inside.  North is top and east is
left. The color in insert (d) shows the OH-maser velocity map,
superposed  on the  L'-band ISAAC
contour image. Note that the scale in (d) is 16$\times$20~arcsec$^2$,
which is different from those in
(a, b and c). }
\label{Fig-NACO}
\end{figure*}

Fig.~\ref{Fig-NACO} shows NACO NB2.12 and L'-band images of OH~231.
The central region is still obscured in NB2.12 image and the bipolar
outflow is brighter.  On the other hand, in L'-band the central region
is brighter than the outflow.  This shows that this central region has
a very red color, because this region is obscured by extremely high
optical depth and/or because it has a very cold temperature, as
suggested by previous infrared observations \citep{Kastner92,
Bujarrabal02}.

In the NB2.12-band image, the central region consists of patchy clouds (blobs)
which were also seen in the near-infrared HST images \citep{Bujarrabal02,
Meakin03}.  
A smooth $\sim
1\times$1~arcsec$^2$ trapezium-shape cloud with a bright central core
is found in the L'-band image (Fig.~\ref{Fig-NACO} (c)).  
This trapezium has
an elongation towards the north-west direction.  This shape
resembles images at 11.7 and 17.9~$\mu$m obtained with the Keck~I
 \citep{Jura02}.

Inside the trapezium cloud in the L'-band image, there is a bright
point-like source.  The NACO observations were unable to resolve the
source, 
suggesting the source size $<0.20\times 0.17$~arcsec$^2$ (FWHM).  
This provides a more stringent limit than
\citet{Jura02}, who reported an unresolved source with a dimension of
$\sim 1\times1$arcsec$^2$ at 11$\mu$m.  
We also our MIDI single-dish acquisition images at
8.7$\mu$m using the AO system (about 0.25~arcsec resolution) also did not
resolve the bright point-like source.

\begin{figure}
\centering
%\resizebox{\hsize}{!}{\includegraphics*{spec_total.ps}}
\resizebox{\hsize}{!}{\includegraphics*{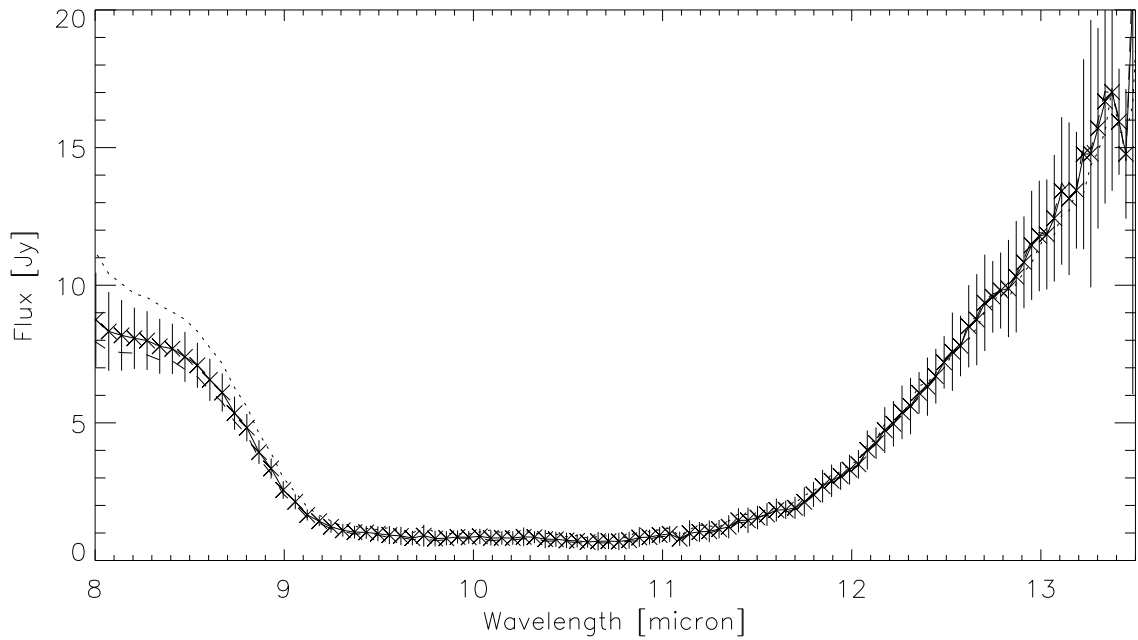}}
\caption{Spectra of OH~231. 
Bold line shows the total spectrum using OB1--OB4 data together 
with the error bars. 
Dotted line shows the spectrum reduced from OB1 (calibrator: HD~50778), 
and the dash line from OB2, OB3 and OB4 (calibrator: HD~61935). 
The difference below 8.5~$\mu$m is probably caused by the uncertainty of 
SiO-band intensities in the spectra of the calibrators. 
}
\label{Fig-spec}
\end{figure}

Fig.~\ref{Fig-spec} shows mid-infrared spectra of OH~231.8+4.2,
obtained with MIDI. This spectrum shows the  flux detected from the
compact source only.
There is a strong
silicate absorption from 9--11~$\mu$m, however, the flux is not completely
zero and is $\sim$1~Jy.  The silicate absorption band is broad for this source.
\citet{Gillett76} and IRAS LRS
observations show 43 and 30~Jy at 8~$\mu$m, respectively and
\citet{Meixner99} reported 25 Jy at 8.8~$\mu$m, while our flux is 
9~Jy at 8~$\mu$m.  Our spectra are from
small core region (about 0.5$\times$0.3~arcsec$^2$), whereas other measurements
for this object
are from more extended regions, up to 2$\times$5~arcsec$^2$ \citep{Jura02}.
Variability at mid-infrared is known for this object \citep{Jura02}
but is not the major reason of the flux discrepancy.

\begin{figure}
\centering
%\resizebox{\hsize}{!}{\includegraphics*{correlated_flux.eps}}
\resizebox{\hsize}{!}{\includegraphics*{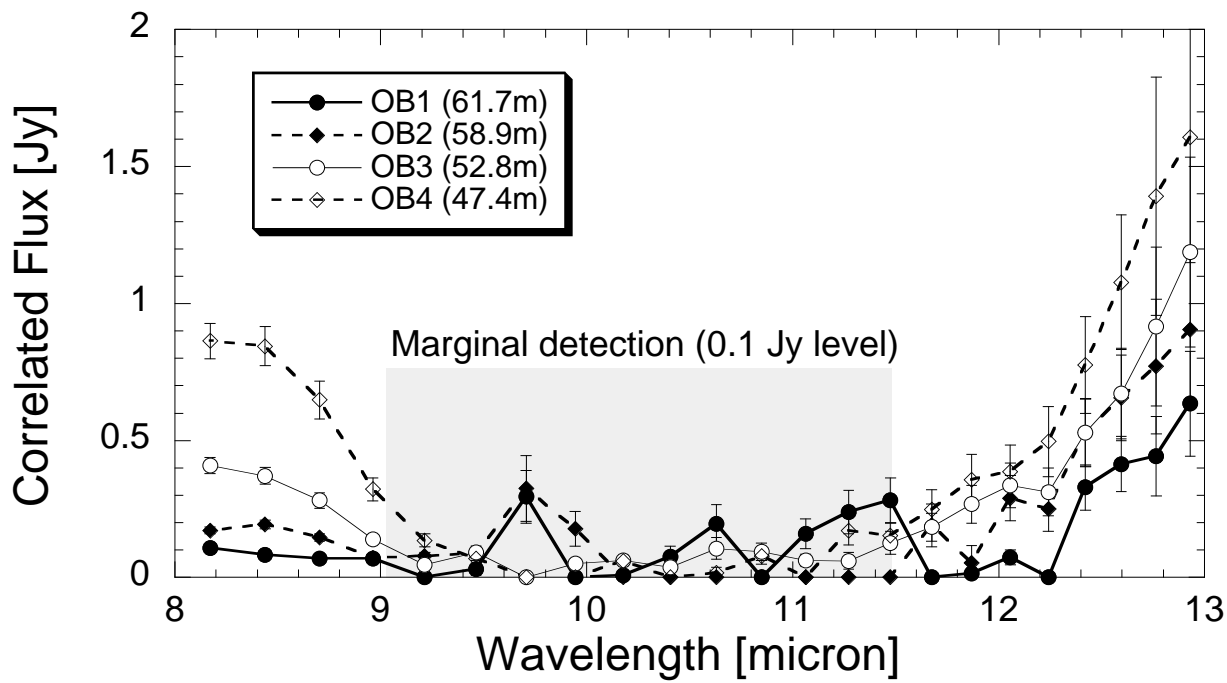}}
\caption{Correlated flux of the mid-infrared `core'.
}
\label{Fig-correlated-flux}
\end{figure}

We detected fringes with MIDI from `the red and point-like source'
found in the near- and mid-infrared images.
Fig.~\ref{Fig-correlated-flux} shows the correlated flux obtained with
MIDI.  Correlated flux is clearly recorded below 9~$\mu$m and above
11~$\mu$m, but there is only a marginal detection ($\sim$0.1~mJy) at
9--11~$\mu$m.  At the wavelength covering the silicate absorption, the
visibilities are small, either because the flux level from this infrared
source is too low or because the source is extended at this wavelength
range.

The visibility data are interpreted in the frame of a smooth Gaussian
profile, as described by \citet{Leinert04}.  The assumption of a
Gaussian profile is appropriate if the object is optically thin
or mildly optically thick.  Fig.~\ref{Fig-gauss} shows radii for the
Gaussian (given in terms of the `half' width of the half maximum; HWHM) required to reproduce the
observed visibilities at each wavelength. We clearly see that the
structure is resolved on the 30--45 mas scale.  At the distance of
1.3~kpc \citep{Jura02}, the radii are about 40--50~AU, as displayed on
the right side of the y-axis in Fig.~\ref{Fig-gauss}.

The position angles of these four baselines vary by only 15 degrees.
The observed differences in the correlated flux between the four
baselines as seen in Fig.~\ref{Fig-correlated-flux} are mainly due to
the baseline length, rather than any asymmetry in the object.  This is
confirmed by the absence of a significant dependence of the 
Gaussian radii on the baseline lengths at each wavelength point 
(Fig.~\ref{Fig-gauss}). These correlated fluxes are therefore resolving 
structures on scales of 60--80 mas in FWHM.

Fig.~\ref{Fig-NACO} (d) presents the velocity distribution
of the 1667~MHz maser emission in m\,s$^{-1}$ 
detected by MERLIN. 
The OH maser data are overplotted on the ISAAC/VLT L'-band image
%(this makes use of the excellent astrometry of the MERLIN and VLT data).
It shows a clear velocity gradient along 
the torus-like maser distribution. Only blue shifted velocities 
were detected by MERLIN.

\section{ Discussion}

Our L'-band and N-band images
using the AO system find an unresolved compact object in the center
with  dimension less
than 200~mas.  This object has a red color and consists of
circumstellar material.  This circumstellar material is responsible
for the fringes detected by mid-infrared interferometer.

\begin{figure}
\centering
%\resizebox{\hsize}{!}{\includegraphics*{fwhm.eps}}
\resizebox{\hsize}{!}{\includegraphics*{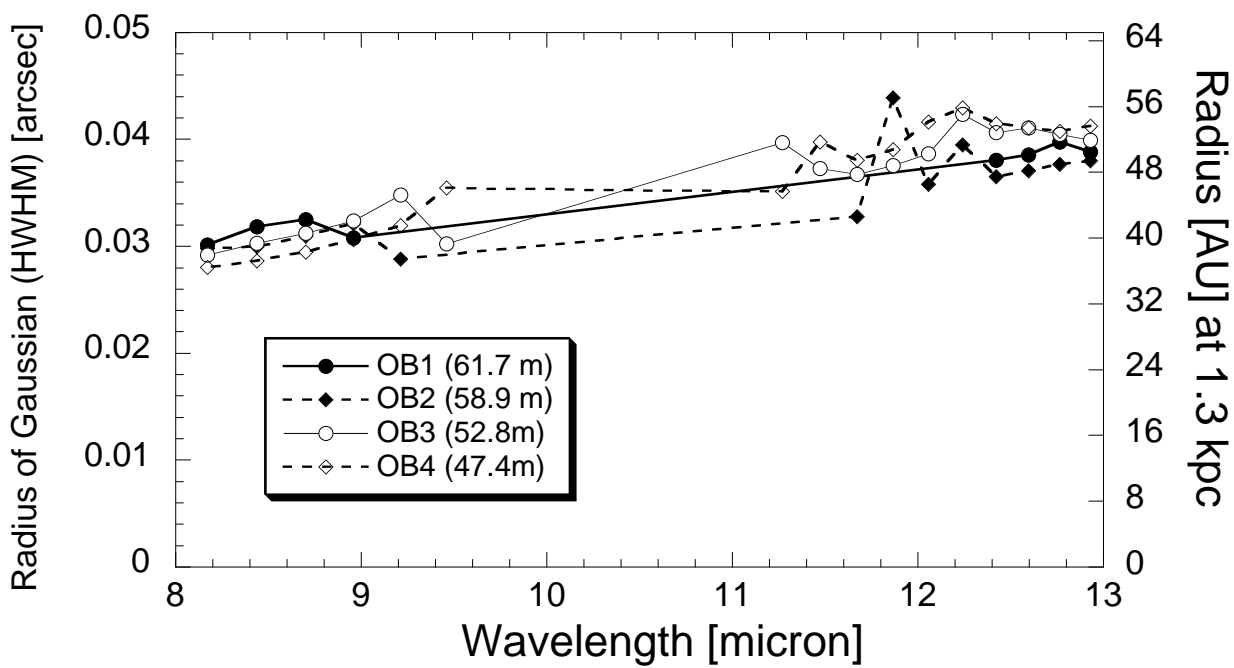}}
\caption{The radius (half of the FWHM of the Gaussian) to reproduce 
the visibilities.
}
\label{Fig-gauss}
\end{figure}

There are other emission mechanisms which can cause correlated flux in
the mid-infrared, such as (1) the central star and (2) the central
star + binary companion(s), but we can exclude these possibilities.
First, the correlated flux of a single star with a radius 3~AU
\citep{SanchezContreras02} at 1.3~kpc (2.3~mas) is almost 100~\% of
the total flux (i.e. unresolved) with current baseline lengths. The
observed fringes indicate a much larger source size.  Using the
parameters of the central star from \citet{Jura02} ($T_{\rm
eff}=2500$~K, radius of $4.6\times10^{13}$~cm), the flux of the
central star is $\sim$6~Jy at 10~$\mu$m, thus the
correlated flux should also be $\sim$6~Jy on any baseline.  The
measured values are inconsistent with the expected correlated flux;
the flux drops to 0.1--0.2~Jy for the longest baseline. 
Thus, the measured visibilities are not from the central star.  
Second, the silicate absorption is detected in correlated flux,
which shows the circumstellar origin of the mid-infrared compact
source.  Although the presence of a binary companion has been
suggested by \citet{SanchezContreras04}, it is unlikely that
we measured the orbit of the binary companion around the central star,
nor the radius of the binary companion.  A sharp angular dependence
of visibilities would be expected in the orbit case, which is not
detected (Fig.~\ref{Fig-gauss}).  The possibility of the binary
companion radius is ruled out because the flux
would be 1~mJy level at 1.3~kpc if the companion is an A0V star
\citep{SanchezContreras04}.  Therefore, the MIDI visibilities are due to
dusty circumstellar material. The absence of stellar emission in
the correlated flux suggests this material is still optically thick at
10~$\mu$m.

\citet{Jura02} analyzed the spectral energy distribution of this
object and argued that the mid-infrared unresolved source is a disk.
\citet{SanchezContreras02} measured the
velocity distribution of SiO maser lines. The SiO masers show
the presence of a rotating disk, within 4~mas from the star. 
\citet{Zijlstra01} find that OH velocity structure could be viewed as an
expanding torus, with an additional component from a bipolar,
ballistic outflow.

Our NACO images also suggest the presence of a disk or torus.  The
central region in the L'-band shows a trapezium shape: this may be
interpreted as a flared outflow from a torus or a disk.  The trapezium
shape is also seen in the NB2.12 image.  The small blobs
seen in the trapezium cloud of NB2.12 image, is probably due to 
non-uniform extinction within the flared disk, or an illusion caused by
scattered light.  The north part of the `trapezium region' is
brighter than other regions in NB2.12, possibly because the disk is
slightly inclined with the northern part nearer to the earth.

The obtained size of the dense circumstellar material is at least
40--50 AU.  The optical depth from our preliminary analysis using 
{\it DUSTY} \citep{Ivezic97} suggests $\tau_{\rm{8\mu m}}$=1.6 and $\tau_{\rm{13~\mu
m}}$=2.7, showing the actual inner radius is smaller than the values
which we measured.  Nevertheless, if the density distribution is
$\rho=\rho_0 (r/r_0)^{-\alpha}$ where $\alpha=1$--2, the optical depth
increases dramatically for closer inner radii, and the
measured size need not be that different from the
actual inner radius of the dusty shell or torus.  The hydrodynamic
model of \citet{Mastro99} shows that an accretion disk with a radius of
40--50 AU can be formed in a binary disk.  The co-incidence of the
radii from the theory and the observations implies that the circumstellar
material is shaped as a disk or torus in this object.

On the other hand, \citet{Jura02} assumed a binary companion with an
orbit of 3--~5~AU for OH 231, and the inner radius of the disk might be
$\sim 1.7$ larger than the companion orbit, which should be 5--9~AU.
Our measured inner radius is much larger than \citet{Jura02}'s
assumption.  This may be because the companion is actually further out
than expected because the central star is larger than expected by
\citet{Jura02}, or because our mid-infrared measurements observe the
radius at which disk has cooled down enough to allow dust to
condense out of the gas phase.
An alternative
solution could be that the disk is gradually expanding and losing
momentum, and at the time of the formation, the disk could be much smaller
than the current size.  In conclusion, we measured the angular size of the
circumstellar material in OH~231to be 40--50~AU, and this material is
probably in a disk or toroidal configuration.  Future MIDI observations with
different baseline angles are required to confirm the presence of a
disk like structure and to measure its inclination angle.

The OH masers are located to the south and east, within the darker lane seen
in the L'-band image, and trace the waist of the trapezium, allowing
for orientation effects.  Only foreground (blue-shifted) OH is seen:
the red emission appears to be more extended and is not picked up by
the MERLIN interferometer, while the blue emission consist of more
compact components. 
The velocity field shows a clear gradient in the southern part of 
the maser spots,
with $v_{\rm LRS}$=20--30\,km\,s$^{-1}$ on the east and west edge and 
the bluest component ($v_{\rm LRS}$=10\,km\,s$^{-1}$) in the middle.
It is inconsistent with an
expanding torus, but is consistent if we observe the blue-shifted rim
of a biconical outflow tilted towards us. Thus, the dynamics of the
source change from a rotating disk in the very centre, within a few
stellar radii, to a conical outflow at $\sim$1000~AU. The stellar
position estimated by the 
the SiO maser tracing an equatorial rotating torus
\citep{SanchezContreras02} coincides with the central position of the
blue-shifted ($\sim$0--5\,km\,s$^{-1}$) biconical distribution seen in OH.

All the components of the binary disk hypothesis \citep{Balick02,
VanWinckel03} may therefore be present in OH~231.8+4.2: a rotating SiO maser disk
very close to the central star, a compact circumstellar material at $\sim$40--50~AU
which may have a disk-like distribution, and a bipolar outflow.

\acknowledgements
We appreciate the ESO staff for their support.
M.M. and E.L. are  supported by PPARC, and S.E. by ESO.
M.M. acknowledges hospitality at
Observatoire de la C\^{o}te d'Azur.

%_________________________________________________________________
\begin{table}
\begin{caption}
{Observing log for MIDI. (Size: the theoretical size of the calibration stars.)
}\label{table-log}
\end{caption}
\begin{tabular}{llcccc}\hline\hline 
OB   & UT time  & Airmass & \multicolumn{2}{c}{Projected baseline} & size \\
 & & & Length  & PA  & \\
 &&&[meters] & [degrees] & [mas] \\
\hline
OB1 (OH~231) & 02:04--02:07 & 1.02 & 61.7 & 112.0  \\
OB2 (OH~231) & 02:53--02:55 & 1.06 & 58.9 & 116.2 \\
OB3 (OH~231) & 03:55--03:57 & 1.19 & 52.8 & 123.8  \\
OB4 (OH~231) & 04:38--04:40 & 1.35 & 47.4 & 131.2  \\
\hline
\multicolumn{2}{l}{Calibrations} \\
Cal1 (HD~50778) & 02:29--02:31 & 1.11 & 56.2 & 117.8 &  3.76$\pm$0.04\\
Cal2 (HD~61935) & 03:12--03:14 & 1.12 & 56.0 & 116.4 &  2.21$\pm$0.01\\
Cal3 (HD~61935) & 04:14--04:16 & 1.30 & 48.3 & 123.8 \\  
\hline
\end{tabular}
\end{table}
%________________________________________________________________

%_________________________________________________________________
%_________________________________________________________________

%_________________________________________________________________
%_________________________________________________________________

%_________________________________________________________________
%_________________________________________________________________

%_________________________________________________________________
%_________________________________________________________________


\begin{thebibliography}{}
\bibliographystyle{aa}

\bibitem[Balick \& Frank(2002)]{Balick02}
 Balick, B., \& Frank, A.,
 2002, ARA\&A 40, 439

\bibitem[Bujarrabal et al.(2002)]{Bujarrabal02}
 Bujarrabal, V., Alcolea, J., S\'{a}nchez-Contreras, C., \& Sahai, R., 
 2002, A\&A 389, 271, 2002

\bibitem[Chesneau et al.(2005)]{Chesneau05a}
 Chesneau, O., Min, M., Herbst, T., et al., 2005, A\&A 435, 1043
 
 \bibitem[Cohen(1981)]{Cohen81}
 Cohen, M.,
 1981, PASP 93, 288
 
\bibitem[Cohen et al.(1985)]{Cohen85}
 Cohen M., Dopita M.A., Schwartz R.D., \& Tielens A.G.G.M.,
 1985, ApJ 287, 702
 
\bibitem[Cohen et al.(1999)]{Cohen99}
 Cohen, M., Walker R.G., Carter B., Hammersley P., Kidger M., \& Noguchi K.,
 1999, AJ 117, 1864
 
\bibitem[Gillett \& Soifer(1976)]{Gillett76}
 Gillett F.C., \& Soifer B.T., 1976, ApJ 207, 780

\bibitem[G\'{o}mez \& Rodr\'{i}guez(2001)]{Gomez01}
 G\'{o}mez, Y., \& Rodr\'{i}guez, L.F.
 2001, ApJ 557, L109

\bibitem[Ivezi\'{c} \& Elitzur(1997)]{Ivezic97}
  Ivezi\'{c}, Z., \& Elitzur, M., 
  1997, MNRAS 287, 799

\bibitem[Jaffe(2004)]{Jaffe04}
  Jaffe, W.J., 
  2004, SPIE 5491, 715

\bibitem[Jura et al.(2002)Jura, Chen, \& Plavchan]{Jura02}
 Jura, M., Chen, C., \& Plavchan, P.,
 2002, ApJ 574, 963

\bibitem[Jura et al.(1985)]{Jura85}
  Jura, M., \& Morris, M., 1985, ApJ, 292, 487

\bibitem[Kastner et al.(1992)]{Kastner92}
  Kastner, J.H., Weintraub, D.A., Zuckerman, B., Becklin, E.E., McLean, I., 
  \& Gatley, I., 1992, ApJ, 398, 552

\bibitem[Leinert et al.(2003)]{Leinert03}
  Leinert Ch., et al., 2003, Msngr, 112, 13

\bibitem[Leinert et al.(2004)]{Leinert04}
  Leinert Ch., et al., 2004, A\&A 423, 537

\bibitem[Lenzen et al.(2003)]{Lenzen03}
 Lenzen R., et al. 2003, SPIE 4841, 944 

\bibitem[Mastrodemos \& Morris(1998)]{Mastro98}
 Mastrodemos N., \& Morris M.,
 1998 ApJ 497, 303

\bibitem[Mastrodemos \& Morris(1999)]{Mastro99}
 Mastrodemos N., \& Morris M.,
 1999 ApJ 523, 357

\bibitem[Meakin et al.(2003)]{Meakin03}
 Meakin C.A., Bieging J.H., Latter W.B., Hora J.L., \& Tielens A.G.G.M.,
 2003 ApJ 585, 482

\bibitem[Meixner et al.(1999)]{Meixner99}
  Meixner M., Ueta T., \& Dayal A.,
 1999, ApJS 122, 221

\bibitem[Rousset et al.(2003)]{Rousset03}
  Rousset, G., et al. 2003, SPIE 4839, 140

\bibitem[S\'{a}nchez\,Contreras et al.(2002)]{SanchezContreras02}
 S\'{a}nchez Contreras C., 
 Desmurs J.F., Bujarrabal J., Alcolea J., \& Colomer F.,
 2002, A\&A 385, L1

\bibitem[S\'{a}nchez\,Contreras et al.(2004)]{SanchezContreras04}
 S\'{a}nchez Contreras C., de Paz, G., \& Sahai, R., 2004, ApJ 616, 519

\bibitem[Van Winckel(2003)]{VanWinckel03}
  Van Winckel, H., 2003, ARA\&A, 41, 391

\bibitem[Zijlstra et al.(2001)]{Zijlstra01}
 Zijlstra A.A., et al.,
 2001, MNRAS 322, 280

\end{thebibliography}
\end{document}